\documentstyle[aps,prl,multicol,epsf]{revtex}

\title{General implementation of all possible positive-operator-value measurements \\ of single photon polarization states} 
 
\author{S. E. Ahnert, M. C. Payne}
\address{Theory of Condensed Matter Group, Cavendish Laboratory, \\
Madingley Road, Cambridge CB3 0HE, U.K.}

\begin{document}
\maketitle 
\begin{abstract}Positive Operator Value Measures (POVMs) are the most general class of quantum measurements. We propose a setup in which all possible POVMs of a single photon polarization state (corresponding to all possible sets of two-dimensional Kraus operators) can be implemented easily using linear optics elements. This method makes it possible to experimentally realize any projective orthogonal, projective non-orthogonal or non-projective sets of any number of POVM operators. Furthermore our implementation only requires vacuum ancillas, and is deterministic rather than probabilistic. Thus it realizes every POVM with the correct set of output states. We give the settings required to implement two different well-known non-orthogonal projective POVMs. 

\bigskip

\end{abstract}

\begin{multicols}{2}

\section{introduction}
The rapidly increasing interest in Quantum Information Theory and its applications - a comprehensive overview of which can be found in \cite{NielsenChuang} - has also generated significant interest in the theory and possible implementations of generalized measurement in the form of Positive Operator Value Measures (POVMs) \cite{Peres,Reznik,Calsamiglia,us}. Such measurements are particularly useful in the context of quantum cryptography \cite{peres,ekert,brandt97,brandt99}. 

Experimentally, a wide variety of quantum mechanical phenomena such as Teleportation \cite{Bennett}, Interaction-Free Measurement \cite{EV} and non-locality \cite{Bell,GHZ,Hardy} have been demonstrated experimentally using photons \cite{Aspect,Zeilinger97,Kwiat99,Zeilinger2000}. Recently it was shown that the operations necessary for quantum computation can be implemented using linear optics \cite{MilburnKnillLaflamme,newmilburn}, which makes photons a promising candidate for quantum information applications. 

We propose here a single setup for the implementation of all possible POVMs of a single-photon polarization state. This includes POVMs with orthogonal projective, non-orthogonal projective, and non-projective sets of Kraus operators. Similar to our previous work in \cite{us}, our setup is a deterministic as opposed to probabilistic implementation of a POVM, which means that the setup delivers one of the possible POVM output states of equation (\ref{povmstat}) in every measurement. In contrast to \cite{us} however, the setup introduced here is much more general and at the same time far simpler. Furthermore our method does not require any ancillas except for vacuum states. Finally, we illustrate how our setup can be used to implement two well-known POVMs, including one required for established quantum cryptography protocols \cite{peres,ekert,brandt97,brandt99}. 

\section{Positive Operator Value Measures}
The positive operator value measure (POVM) is the most general formulation of quantum measurement \cite{bibkraus}. Mathematically it corresponds to a positive-definite partition of unity in the space of operators on a given Hilbert space. A POVM is given by a set of positive definite Hermitian operators $\{F_i\}$, which in turn can be expressed in terms a set of so-called Kraus operators $\{M_i\}$, such that $F_i = M^\dagger_i M_i$ and for a POVM with $n$ operators,  

\begin{equation}
\sum_{i=1}^n M_i^\dagger M_i = \sum_{i=1}^n F_i = I
\end{equation}

where $I$ is the unit matrix.
 After a POVM measurement is performed on a quantum state represented by a density matrix $\rho$, the state becomes $\rho'$, where

\begin{equation}\label{povmstat}
\rho' = {M_i \rho M_i^\dagger \over {\rm tr}(M_i \rho M_i^\dagger)}
\end{equation}

with probability $p_i$, where 

\begin{equation}\label{povmprob}
p_i = {\rm tr}(M_i \rho M_i^\dagger)
\end{equation}

Note that a POVM can project the input state to a fixed set of (orthogonal or non-orthogonal) states, but can also be non-projective, meaning that the set of possible final states is not fixed, but depends on the input state.
In the case of a set of projective orthogonal operators all $M_i$ can be written as outer products of pairs of orthogonal state vectors. For projective non-orthogonal operators all $M_i$ can be written as outer products of general state vectors. Finally, non-projective operators are all remaining sets, i.e. sets which contain at least one member which cannot be written as an outer product of state vectors. 

 A deterministic implementation (such as the setup we present here) of any of these types of POVM gives one of the possible POVM output states of eq. (\ref{povmstat}) in every measurement, with probabilities of eq. (\ref{povmprob}). A probabilistic implementation by contrast would only give the probability distribution of eq. (\ref{povmprob}), and may not even be successful in performing the POVM every time.   

\section{the povm operator module}

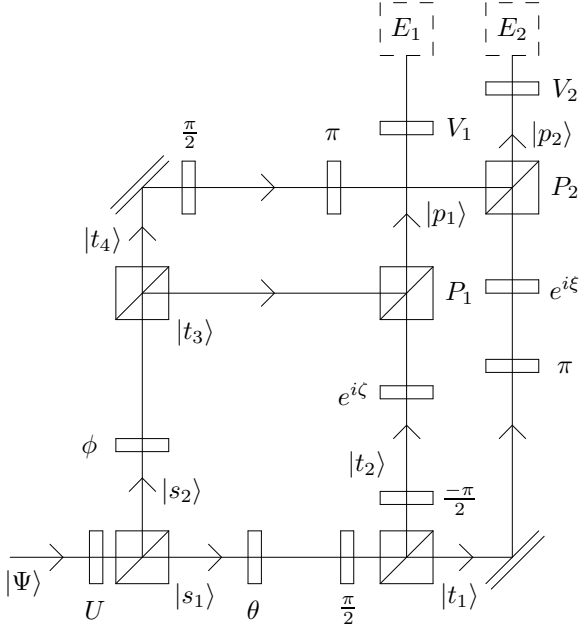
\begin{figure}

\setlength{\unitlength}{1pt}
\ifx\plotpoint\undefined\newsavebox{\plotpoint}\fi
\begin{picture}(250,290)(0,0)
\font\gnuplot=cmr10 at 10pt
\gnuplot
\put(50,50){\special{em:moveto}}
\put(70,50){\special{em:lineto}}
\put(70,70){\special{em:lineto}}
\put(50,70){\special{em:lineto}}
\put(50,50){\special{em:lineto}}
\put(70,70){\special{em:lineto}}
\put(190,190){\special{em:moveto}}
\put(210,190){\special{em:lineto}}
\put(210,210){\special{em:lineto}}
\put(190,210){\special{em:lineto}}
\put(190,190){\special{em:lineto}}
\put(210,210){\special{em:lineto}}
\put(150,150){\special{em:moveto}}
\put(170,150){\special{em:lineto}}
\put(170,170){\special{em:lineto}}
\put(150,170){\special{em:lineto}}
\put(150,150){\special{em:lineto}}
\put(170,170){\special{em:lineto}}
\put(150,50){\special{em:moveto}}
\put(170,50){\special{em:lineto}}
\put(170,70){\special{em:lineto}}
\put(150,70){\special{em:lineto}}
\put(150,50){\special{em:lineto}}
\put(170,70){\special{em:lineto}}
\put(50,150){\special{em:moveto}}
\put(70,150){\special{em:lineto}}
\put(70,170){\special{em:lineto}}
\put(50,170){\special{em:lineto}}
\put(50,150){\special{em:lineto}}
\put(70,170){\special{em:lineto}}
\put(10,60){\special{em:moveto}}
\put(200,60){\special{em:lineto}}
\put(200,250){\special{em:lineto}}
\put(60,60){\special{em:moveto}}
\put(60,200){\special{em:lineto}}
\put(200,200){\special{em:lineto}}
\put(60,160){\special{em:moveto}}
\put(160,160){\special{em:lineto}}
\put(160,60){\special{em:moveto}}
\put(160,250){\special{em:lineto}}
\put(190,50){\special{em:moveto}}
\put(210,70){\special{em:lineto}}
\put(192,48){\special{em:moveto}}
\put(212,68){\special{em:lineto}}
\put(50,190){\special{em:moveto}}
\put(70,210){\special{em:lineto}}
\put(48,192){\special{em:moveto}}
\put(68,212){\special{em:lineto}}

\put(135,50){\special{em:moveto}}
\put(140,50){\special{em:lineto}}
\put(140,70){\special{em:lineto}}
\put(135,70){\special{em:lineto}}
\put(135,50){\special{em:lineto}}
\put(100,50){\special{em:moveto}}
\put(105,50){\special{em:lineto}}
\put(105,70){\special{em:lineto}}
\put(100,70){\special{em:lineto}}
\put(100,50){\special{em:lineto}}
\put(80,190){\special{em:moveto}}
\put(75,190){\special{em:lineto}}
\put(75,210){\special{em:lineto}}
\put(80,210){\special{em:lineto}}
\put(80,190){\special{em:lineto}}
\put(150,80){\special{em:moveto}}
\put(170,80){\special{em:lineto}}
\put(170,85){\special{em:lineto}}
\put(150,85){\special{em:lineto}}
\put(150,80){\special{em:lineto}}
\put(190,130){\special{em:moveto}}
\put(210,130){\special{em:lineto}}
\put(210,135){\special{em:lineto}}
\put(190,135){\special{em:lineto}}
\put(190,130){\special{em:lineto}}
\put(130,190){\special{em:moveto}}
\put(130,210){\special{em:lineto}}
\put(135,210){\special{em:lineto}}
\put(135,190){\special{em:lineto}}
\put(130,190){\special{em:lineto}}
\put(50,100){\special{em:moveto}}
\put(70,100){\special{em:lineto}}
\put(70,105){\special{em:lineto}}
\put(50,105){\special{em:lineto}}
\put(50,100){\special{em:lineto}}
\put(25,55){\special{em:moveto}}
\put(30,60){\special{em:lineto}}
\put(25,65){\special{em:lineto}}
\put(155,185){\special{em:moveto}}
\put(160,190){\special{em:lineto}}
\put(165,185){\special{em:lineto}}
\put(195,215){\special{em:moveto}}
\put(200,220){\special{em:lineto}}
\put(205,215){\special{em:lineto}}
\put(85,55){\special{em:moveto}}
\put(90,60){\special{em:lineto}}
\put(85,65){\special{em:lineto}}
\put(155,105){\special{em:moveto}}
\put(160,110){\special{em:lineto}}
\put(165,105){\special{em:lineto}}
\put(105,155){\special{em:moveto}}
\put(110,160){\special{em:lineto}}
\put(105,165){\special{em:lineto}}
\put(55,85){\special{em:moveto}}
\put(60,90){\special{em:lineto}}
\put(65,85){\special{em:lineto}}
\put(105,195){\special{em:moveto}}
\put(110,200){\special{em:lineto}}
\put(105,205){\special{em:lineto}}
\put(180,55){\special{em:moveto}}
\put(185,60){\special{em:lineto}}
\put(180,65){\special{em:lineto}}
\put(55,180){\special{em:moveto}}
\put(60,185){\special{em:lineto}}
\put(65,180){\special{em:lineto}}
\put(195,105){\special{em:moveto}}
\put(200,110){\special{em:lineto}}
\put(205,105){\special{em:lineto}}
\put(190,250){\special{em:moveto}}
\put(195,250){\special{em:lineto}}
\put(200,250){\special{em:moveto}}
\put(205,250){\special{em:lineto}}
\put(210,250){\special{em:moveto}}
\put(210,255){\special{em:lineto}}
\put(210,260){\special{em:moveto}}
\put(210,265){\special{em:lineto}}
\put(210,270){\special{em:moveto}}
\put(205,270){\special{em:lineto}}
\put(200,270){\special{em:moveto}}
\put(195,270){\special{em:lineto}}
\put(190,270){\special{em:moveto}}
\put(190,265){\special{em:lineto}}
\put(190,260){\special{em:moveto}}
\put(190,255){\special{em:lineto}}
\put(150,250){\special{em:moveto}}
\put(155,250){\special{em:lineto}}
\put(160,250){\special{em:moveto}}
\put(165,250){\special{em:lineto}}
\put(170,250){\special{em:moveto}}
\put(170,255){\special{em:lineto}}
\put(170,260){\special{em:moveto}}
\put(170,265){\special{em:lineto}}
\put(170,270){\special{em:moveto}}
\put(165,270){\special{em:lineto}}
\put(160,270){\special{em:moveto}}
\put(155,270){\special{em:lineto}}
\put(150,270){\special{em:moveto}}
\put(150,265){\special{em:lineto}}
\put(150,260){\special{em:moveto}}
\put(150,255){\special{em:lineto}}
\put(150,120){\special{em:moveto}}
\put(170,120){\special{em:lineto}}
\put(170,125){\special{em:lineto}}
\put(150,125){\special{em:lineto}}
\put(150,120){\special{em:lineto}}
\put(190,160){\special{em:moveto}}
\put(210,160){\special{em:lineto}}
\put(210,165){\special{em:lineto}}
\put(190,165){\special{em:lineto}}
\put(190,160){\special{em:lineto}}
\put(150,220){\special{em:moveto}}
\put(170,220){\special{em:lineto}}
\put(170,225){\special{em:lineto}}
\put(150,225){\special{em:lineto}}
\put(150,220){\special{em:lineto}}
\put(190,235){\special{em:moveto}}
\put(210,235){\special{em:lineto}}
\put(210,240){\special{em:lineto}}
\put(190,240){\special{em:lineto}}
\put(190,235){\special{em:lineto}}
\put(45,50){\special{em:moveto}}
\put(40,50){\special{em:lineto}}
\put(40,70){\special{em:lineto}}
\put(45,70){\special{em:lineto}}
\put(45,50){\special{em:lineto}}

\put(102,40){\makebox(0,0){$\theta$}}
\put(132,220){\makebox(0,0){$\pi$}}
\put(40,102){\makebox(0,0){$\phi$}}
\put(138,40){\makebox(0,0){$\pi \over 2$}}
\put(78,220){\makebox(0,0){$\pi \over 2$}}
\put(180,82){\makebox(0,0){$-\pi \over 2$}}
\put(220,132){\makebox(0,0){$\pi$}}
\put(220,200){\makebox(0,0){$P_2$}}
\put(180,160){\makebox(0,0){$P_1$}}
\put(215,220){\makebox(0,0){$| p_2 \rangle$}}
\put(175,190){\makebox(0,0){$| p_1 \rangle$}}
\put(75,85){\makebox(0,0){$| s_2 \rangle$}}
\put(80,45){\makebox(0,0){$| s_1 \rangle$}}
\put(180,45){\makebox(0,0){$| t_1 \rangle$}}
\put(145,95){\makebox(0,0){$| t_2 \rangle$}}
\put(80,145){\makebox(0,0){$| t_3 \rangle$}}
\put(45,180){\makebox(0,0){$| t_4 \rangle$}}
\put(140,122){\makebox(0,0){$e^{i\zeta}$}}
\put(220,162){\makebox(0,0){$e^{i\xi}$}}
\put(42,40){\makebox(0,0){$U$}}
\put(180,222){\makebox(0,0){$V_1$}}
\put(220,237){\makebox(0,0){$V_2$}}
\put(15,50){\makebox(0,0){$| \Psi \rangle$}}
\put(160,260){\makebox(0,0){$E_1$}}
\put(200,260){\makebox(0,0){$E_2$}}
\end{picture}

\vspace{-1cm}

\caption{The module implementing measurement operators $F_1$ and $F_2$. The photon enters in state $| \Psi \rangle$ at the bottom left corner and exits either at $E_1$ or $E_2$, where it can be detected. All beamsplitters are polarizing beamsplitters with the same polarization basis and transmit photons in the $| H \rangle$ state, while reflecting photons in the $| V \rangle$ state. The angles $\theta$, $\phi$, $\pi \over 2$ and $\pi$ of the polarization rotators are measured relative to this basis. $U$, $V_1$ and $V_2$ are unitary operators, and $e^{i \zeta}$ and $e^{i \xi}$ signify phase shifters.}\label{module}
\end{figure}

The implementation proposed here uses $(n-1)$ linear optics modules for an $n$ operator POVM. One such module is depicted in Fig. \ref{module}. 
It consists of five polarizing beamsplitters, arranged as shown in this figure. The photon enters at the bottom left of Fig.\ref{module} and is split into its horizontal and vertical polarization states $| H \rangle$ and $| V \rangle$ respectively. These components, now in the path (or 'which-path') states $| s_1 \rangle$ (H component) and $| s_2 \rangle$ (V component) are then rotated by angles $\theta$ (H component) and $\phi$ (V component), using variable polarization rotators. Then both of these amplitudes are in turn split by two further polarizing beamsplitters and form a superposition of the four path states $| t_1 \rangle$ to $| t_4 \rangle$. The beamsplitters $P_1$ then reunifies the path states $| t_2 \rangle$ and $| t_3 \rangle$ in $| p_1 \rangle$, and $P_2$ recombines $| t_1 \rangle$ and $| t_4 \rangle$ in $| p_2 \rangle$. Furthermore the setup contains another five polarization rotators (rotating by angles $+{\pi \over 2}$, $-{\pi \over 2}$ and ${\pi}$), three variable unitary operators $U$, $V_1$ and $V_2$, and two variable phase shifts $e^{i \zeta}$ and $e^{i \xi}$. All these elements are placed as shown in Fig. \ref{module}. Ancillas, often required in linear optics implementations of quantum information processes are here only present in form of vacuum states in the ports of the polarizing beamsplitters.

\vspace{-1cm}

Consider the case where $U = V_1 = V_2 = I$ (the unit matrix) and $\zeta = \xi = 0$. Then a photon incident on the apparatus in the state

\begin{equation}
| \Psi \rangle = a | H \rangle + b | V \rangle
\end{equation}

where $|a|^2 + |b|^2 = 1$, evolves to:

\begin{eqnarray}
| \Psi \rangle &\rightarrow& a | H \rangle | s_1 \rangle + b | V \rangle | s_2 \rangle \cr\cr
&\rightarrow& a (\cos \theta | H \rangle + \sin \theta | V \rangle ) | s_1 \rangle \cr
&&+ b (\cos \phi | V \rangle - \sin \phi | H \rangle ) | s_2 \rangle \cr\cr
&\rightarrow& a (\cos \theta | V \rangle - \sin \theta | H \rangle ) | s_1 \rangle\cr
&&+ b (\cos \phi | V \rangle - \sin \phi | H \rangle ) | s_2 \rangle \cr\cr
&\rightarrow& a (\cos \theta | V \rangle | t_2 \rangle - \sin \theta | H \rangle | t_1 \rangle) \cr
&&+ b (\cos \phi | V \rangle | t_3 \rangle - \sin \phi | H \rangle | t_4 \rangle) \cr\cr
&\rightarrow& a (\cos \theta | H \rangle | t_2 \rangle + \sin \theta | H \rangle | t_1 \rangle) \cr
&&+ b (\cos \phi | V \rangle | t_3 \rangle + \sin \phi | V \rangle | t_4 \rangle) \cr
\end{eqnarray}

The beamsplitter $P_1$ then recombines path states $| t_2 \rangle$ and $| t_3 \rangle$, and similarly $P_2$ recombines $| t_1 \rangle$ and $| t_4 \rangle$, so that: 

\begin{eqnarray}\label{psi1}
| \Psi \rangle \rightarrow (a \cos \theta | H \rangle + b \cos \phi | V)| p_1 \rangle \cr + (a \sin \theta | H \rangle + b \sin \phi | V \rangle)| p_2 \rangle  \end{eqnarray}
 
where $| p_1 \rangle$ and $| p_2 \rangle$ denote the path states of amplitudes emerging from beamsplitter $P_1$ and $P_2$ respectively.

This corresponds to the matrix transformations

\begin{equation}\label{d1}
\left(
\begin{array}{c}
a \cr b
\end{array}
\right)
\rightarrow
D_1
\left(
\begin{array}{c}
a \cr b
\end{array}
\right)
=\left(
\begin{array}{cc}
\cos \theta & 0 \cr 0 & \cos \phi
\end{array}
\right)
\left(
\begin{array}{c}
a \cr b
\end{array}
\right)
\end{equation}

\begin{equation}\label{d2}
\left(
\begin{array}{c}
a \cr b
\end{array}
\right)
\rightarrow
D_2
\left(
\begin{array}{c}
a \cr b
\end{array}
\right)
=
\left(
\begin{array}{cc}
\sin \theta & 0 \cr 0 & \sin \phi 
\end{array}
\right)
\left(
\begin{array}{c}
a \cr b
\end{array}
\right)
\end{equation}

in the respective Hilbert spaces of $| p_1 \rangle$ and $| p_2 \rangle$.
Note that, as required, ${D_1}^2 + {D_2}^2 = I$ where $I$ is the unit matrix. It is due to the vacuum state ancilla entering the first beamsplitter that two of the four outputs of the two final beamsplitters remain dark, giving a partition of unity into two operators rather than four. 

An arbitrary $n \times n$ matrix $A$ can be written as $A=VDU$ where $V$ and $U$ are unitary matrices and $D$ is a diagonal matrix. Thus we can write any general Kraus operator $M$ for quantum measurement as:

\begin{equation}\label{kraus}
M_i = V_i D_i U_i 
\end{equation}

The moduli of the elements of the diagonal matrix $D_i$ are confined to lie between 0 and 1. 
(Note that in general $M_i \neq M_i^{\dagger}$.) 

Let us consider the two operators in our module, with $U_1 = U_2$ implemented by a variable polarization rotator, placed before the entrance of the module. Also we introduce phase shifts $\zeta$ and $\xi$ for the sake of generality, as the $\{D_i\}$ in equation (\ref{kraus}) are in general complex. Hence:

\begin{equation}
D_1=
\left(
\begin{array}{cc}
e^{i\zeta} \cos \theta & 0 \cr 0 & \cos \phi 
\end{array}
\right)
\,\,\,
D_2=
\left(
\begin{array}{cc}
e^{i\xi} \sin \theta & 0 \cr 0 & \sin \phi 
\end{array}
\right)
\end{equation}

Thus:

\begin{equation}
F_1 = M_1^{\dagger} M_1 = U^{\dagger}_1 D_1^2 U_1
\end{equation}

where $D_1^2 \equiv D_1^{\dagger} V_1^{\dagger} V_1 D_1 = D_1^{\dagger} D_1$. Hence

\begin{eqnarray}
F_2 = I - F_1 = I - U^{\dagger}_1 D_1^2 U_1 = U^\dagger_1 U_1 - U^{\dagger}_1 D_1^2 U_1 
\cr 
= U^{\dagger}_1 (I - D_1^2) U_1 = U^{\dagger}_1 D_2^2 U_1
\end{eqnarray}

as required by $\sum_{i=1}^n M_i^\dagger M_i = \sum_{i=1}^n F_i = I$. This also makes it clear that $U_1 = U_2$ follows naturally and does not place an additional constraint on the space of possible operators that can be implemented. Hence the arrangement illustrated in Fig. \ref{module} provides a physical implementation of a completely general positive-definite bipartition of unity, which is what a two-operator POVM represents in mathematical terms. 
This apparatus is deterministic, which means that all output states are given by equation (\ref{povmstat}) with probabilities given in equation (\ref{povmprob}). It therefore does not only implement the POVM operators, but also specific Kraus operators, chosen by the operators $\{V_i\}$ in equation (\ref{kraus}). These operators are implemented at the exits of the operator modules, as shown in Figs. \ref{module} and \ref{twomod}.

\section{Generalization to N modules}
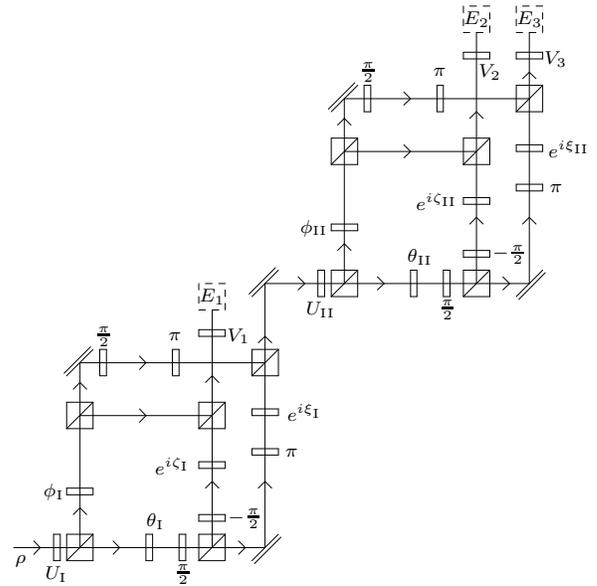
\begin{figure}

\setlength{\unitlength}{0.5pt}
\ifx\plotpoint\undefined\newsavebox{\plotpoint}\fi
\begin{picture}(450,470)(0,0)
\font\gnuplot=cmr10 at 10pt
\gnuplot
\put(50,50){\special{em:moveto}}
\put(70,50){\special{em:lineto}}
\put(70,70){\special{em:lineto}}
\put(50,70){\special{em:lineto}}
\put(50,50){\special{em:lineto}}
\put(70,70){\special{em:lineto}}
\put(190,190){\special{em:moveto}}
\put(210,190){\special{em:lineto}}
\put(210,210){\special{em:lineto}}
\put(190,210){\special{em:lineto}}
\put(190,190){\special{em:lineto}}
\put(210,210){\special{em:lineto}}
\put(150,150){\special{em:moveto}}
\put(170,150){\special{em:lineto}}
\put(170,170){\special{em:lineto}}
\put(150,170){\special{em:lineto}}
\put(150,150){\special{em:lineto}}
\put(170,170){\special{em:lineto}}
\put(150,50){\special{em:moveto}}
\put(170,50){\special{em:lineto}}
\put(170,70){\special{em:lineto}}
\put(150,70){\special{em:lineto}}
\put(150,50){\special{em:lineto}}
\put(170,70){\special{em:lineto}}
\put(50,150){\special{em:moveto}}
\put(70,150){\special{em:lineto}}
\put(70,170){\special{em:lineto}}
\put(50,170){\special{em:lineto}}
\put(50,150){\special{em:lineto}}
\put(70,170){\special{em:lineto}}
\put(10,60){\special{em:moveto}}
\put(200,60){\special{em:lineto}}
\put(200,260){\special{em:lineto}}
\put(60,60){\special{em:moveto}}
\put(60,200){\special{em:lineto}}
\put(200,200){\special{em:lineto}}
\put(60,160){\special{em:moveto}}
\put(160,160){\special{em:lineto}}
\put(160,60){\special{em:moveto}}
\put(160,240){\special{em:lineto}}
\put(190,50){\special{em:moveto}}
\put(210,70){\special{em:lineto}}
\put(192,48){\special{em:moveto}}
\put(212,68){\special{em:lineto}}
\put(50,190){\special{em:moveto}}
\put(70,210){\special{em:lineto}}
\put(48,192){\special{em:moveto}}
\put(68,212){\special{em:lineto}}
\put(110,50){\special{em:moveto}}
\put(115,50){\special{em:lineto}}
\put(115,70){\special{em:lineto}}
\put(110,70){\special{em:lineto}}
\put(110,50){\special{em:lineto}}
\put(40,50){\special{em:moveto}}
\put(45,50){\special{em:lineto}}
\put(45,70){\special{em:lineto}}
\put(40,70){\special{em:lineto}}
\put(40,50){\special{em:lineto}}
\put(150,120){\special{em:moveto}}
\put(170,120){\special{em:lineto}}
\put(170,125){\special{em:lineto}}
\put(150,125){\special{em:lineto}}
\put(150,120){\special{em:lineto}}
\put(190,160){\special{em:moveto}}
\put(210,160){\special{em:lineto}}
\put(210,165){\special{em:lineto}}
\put(190,165){\special{em:lineto}}
\put(190,160){\special{em:lineto}}
\put(135,50){\special{em:moveto}}
\put(140,50){\special{em:lineto}}
\put(140,70){\special{em:lineto}}
\put(135,70){\special{em:lineto}}
\put(135,50){\special{em:lineto}}
\put(80,190){\special{em:moveto}}
\put(75,190){\special{em:lineto}}
\put(75,210){\special{em:lineto}}
\put(80,210){\special{em:lineto}}
\put(80,190){\special{em:lineto}}
\put(150,80){\special{em:moveto}}
\put(170,80){\special{em:lineto}}
\put(170,85){\special{em:lineto}}
\put(150,85){\special{em:lineto}}
\put(150,80){\special{em:lineto}}
\put(190,130){\special{em:moveto}}
\put(210,130){\special{em:lineto}}
\put(210,135){\special{em:lineto}}
\put(190,135){\special{em:lineto}}
\put(190,130){\special{em:lineto}}
\put(50,100){\special{em:moveto}}
\put(70,100){\special{em:lineto}}
\put(70,105){\special{em:lineto}}
\put(50,105){\special{em:lineto}}
\put(50,100){\special{em:lineto}}
\put(25,55){\special{em:moveto}}
\put(30,60){\special{em:lineto}}
\put(25,65){\special{em:lineto}}
\put(155,185){\special{em:moveto}}
\put(160,190){\special{em:lineto}}
\put(165,185){\special{em:lineto}}
\put(195,215){\special{em:moveto}}
\put(200,220){\special{em:lineto}}
\put(205,215){\special{em:lineto}}
\put(85,55){\special{em:moveto}}
\put(90,60){\special{em:lineto}}
\put(85,65){\special{em:lineto}}
\put(155,105){\special{em:moveto}}
\put(160,110){\special{em:lineto}}
\put(165,105){\special{em:lineto}}
\put(105,155){\special{em:moveto}}
\put(110,160){\special{em:lineto}}
\put(105,165){\special{em:lineto}}
\put(55,85){\special{em:moveto}}
\put(60,90){\special{em:lineto}}
\put(65,85){\special{em:lineto}}
\put(105,195){\special{em:moveto}}
\put(110,200){\special{em:lineto}}
\put(105,205){\special{em:lineto}}
\put(180,55){\special{em:moveto}}
\put(185,60){\special{em:lineto}}
\put(180,65){\special{em:lineto}}
\put(55,180){\special{em:moveto}}
\put(60,185){\special{em:lineto}}
\put(65,180){\special{em:lineto}}
\put(195,105){\special{em:moveto}}
\put(200,110){\special{em:lineto}}
\put(205,105){\special{em:lineto}}
\put(130,190){\special{em:moveto}}
\put(130,210){\special{em:lineto}}
\put(135,210){\special{em:lineto}}
\put(135,190){\special{em:lineto}}
\put(130,190){\special{em:lineto}}

\put(188,252){\special{em:moveto}}
\put(208,272){\special{em:lineto}}
\put(190,250){\special{em:moveto}}
\put(210,270){\special{em:lineto}}

\put(150,240){\special{em:moveto}}
\put(155,240){\special{em:lineto}}
\put(160,240){\special{em:moveto}}
\put(165,240){\special{em:lineto}}
\put(170,240){\special{em:moveto}}
\put(170,245){\special{em:lineto}}
\put(170,250){\special{em:moveto}}
\put(170,255){\special{em:lineto}}
\put(170,260){\special{em:moveto}}
\put(165,260){\special{em:lineto}}
\put(160,260){\special{em:moveto}}
\put(155,260){\special{em:lineto}}
\put(150,260){\special{em:moveto}}
\put(150,255){\special{em:lineto}}
\put(150,250){\special{em:moveto}}
\put(150,245){\special{em:lineto}}
\put(117,80){\makebox(0,0){\scriptsize $\theta_{\rm I}$}}
\put(130,122){\makebox(0,0){\scriptsize $e^{i\zeta_{\rm I}}$}}
\put(230,162){\makebox(0,0){\scriptsize $e^{i\xi_{\rm I}}$}}
\put(42,40){\makebox(0,0){\scriptsize $U_{\rm I}$}}
\put(40,102){\makebox(0,0){\scriptsize $\phi_{\rm I}$}}
\put(138,40){\makebox(0,0){\scriptsize $\pi \over 2$}}
\put(78,220){\makebox(0,0){\scriptsize $\pi \over 2$}}
\put(185,82){\makebox(0,0){\scriptsize $-{\pi \over 2}$}}
\put(220,132){\makebox(0,0){\scriptsize $\pi$}}
\put(132,220){\makebox(0,0){\scriptsize $\pi$}}

\put(250,250){\special{em:moveto}}
\put(270,250){\special{em:lineto}}
\put(270,270){\special{em:lineto}}
\put(250,270){\special{em:lineto}}
\put(250,250){\special{em:lineto}}
\put(270,270){\special{em:lineto}}
\put(390,390){\special{em:moveto}}
\put(410,390){\special{em:lineto}}
\put(410,410){\special{em:lineto}}
\put(390,410){\special{em:lineto}}
\put(390,390){\special{em:lineto}}
\put(410,410){\special{em:lineto}}
\put(350,350){\special{em:moveto}}
\put(370,350){\special{em:lineto}}
\put(370,370){\special{em:lineto}}
\put(350,370){\special{em:lineto}}
\put(350,350){\special{em:lineto}}
\put(370,370){\special{em:lineto}}
\put(350,250){\special{em:moveto}}
\put(370,250){\special{em:lineto}}
\put(370,270){\special{em:lineto}}
\put(350,270){\special{em:lineto}}
\put(350,250){\special{em:lineto}}
\put(370,270){\special{em:lineto}}
\put(250,350){\special{em:moveto}}
\put(270,350){\special{em:lineto}}
\put(270,370){\special{em:lineto}}
\put(250,370){\special{em:lineto}}
\put(250,350){\special{em:lineto}}
\put(270,370){\special{em:lineto}}
\put(330,390){\special{em:moveto}}
\put(330,410){\special{em:lineto}}
\put(335,410){\special{em:lineto}}
\put(335,390){\special{em:lineto}}
\put(330,390){\special{em:lineto}}
\put(200,260){\special{em:moveto}}
\put(400,260){\special{em:lineto}}
\put(400,450){\special{em:lineto}}
\put(260,260){\special{em:moveto}}
\put(260,400){\special{em:lineto}}
\put(400,400){\special{em:lineto}}
\put(260,360){\special{em:moveto}}
\put(360,360){\special{em:lineto}}
\put(360,260){\special{em:moveto}}
\put(360,450){\special{em:lineto}}
\put(390,250){\special{em:moveto}}
\put(410,270){\special{em:lineto}}
\put(392,248){\special{em:moveto}}
\put(412,268){\special{em:lineto}}
\put(250,390){\special{em:moveto}}
\put(270,410){\special{em:lineto}}
\put(248,392){\special{em:moveto}}
\put(268,412){\special{em:lineto}}
\put(310,250){\special{em:moveto}}
\put(315,250){\special{em:lineto}}
\put(315,270){\special{em:lineto}}
\put(310,270){\special{em:lineto}}
\put(310,250){\special{em:lineto}}
\put(240,250){\special{em:moveto}}
\put(245,250){\special{em:lineto}}
\put(245,270){\special{em:lineto}}
\put(240,270){\special{em:lineto}}
\put(240,250){\special{em:lineto}}
\put(350,320){\special{em:moveto}}
\put(370,320){\special{em:lineto}}
\put(370,325){\special{em:lineto}}
\put(350,325){\special{em:lineto}}
\put(350,320){\special{em:lineto}}
\put(390,360){\special{em:moveto}}
\put(410,360){\special{em:lineto}}
\put(410,365){\special{em:lineto}}
\put(390,365){\special{em:lineto}}
\put(390,360){\special{em:lineto}}
\put(335,250){\special{em:moveto}}
\put(340,250){\special{em:lineto}}
\put(340,270){\special{em:lineto}}
\put(335,270){\special{em:lineto}}
\put(335,250){\special{em:lineto}}
\put(280,390){\special{em:moveto}}
\put(275,390){\special{em:lineto}}
\put(275,410){\special{em:lineto}}
\put(280,410){\special{em:lineto}}
\put(280,390){\special{em:lineto}}
\put(350,280){\special{em:moveto}}
\put(370,280){\special{em:lineto}}
\put(370,285){\special{em:lineto}}
\put(350,285){\special{em:lineto}}
\put(350,280){\special{em:lineto}}
\put(390,330){\special{em:moveto}}
\put(410,330){\special{em:lineto}}
\put(410,335){\special{em:lineto}}
\put(390,335){\special{em:lineto}}
\put(390,330){\special{em:lineto}}
\put(250,300){\special{em:moveto}}
\put(270,300){\special{em:lineto}}
\put(270,305){\special{em:lineto}}
\put(250,305){\special{em:lineto}}
\put(250,300){\special{em:lineto}}
\put(225,255){\special{em:moveto}}
\put(230,260){\special{em:lineto}}
\put(225,265){\special{em:lineto}}
\put(355,385){\special{em:moveto}}
\put(360,390){\special{em:lineto}}
\put(365,385){\special{em:lineto}}
\put(395,415){\special{em:moveto}}
\put(400,420){\special{em:lineto}}
\put(405,415){\special{em:lineto}}
\put(285,255){\special{em:moveto}}
\put(290,260){\special{em:lineto}}
\put(285,265){\special{em:lineto}}
\put(355,305){\special{em:moveto}}
\put(360,310){\special{em:lineto}}
\put(365,305){\special{em:lineto}}
\put(305,355){\special{em:moveto}}
\put(310,360){\special{em:lineto}}
\put(305,365){\special{em:lineto}}
\put(255,285){\special{em:moveto}}
\put(260,290){\special{em:lineto}}
\put(265,285){\special{em:lineto}}
\put(305,395){\special{em:moveto}}
\put(310,400){\special{em:lineto}}
\put(305,405){\special{em:lineto}}
\put(380,255){\special{em:moveto}}
\put(385,260){\special{em:lineto}}
\put(380,265){\special{em:lineto}}
\put(255,380){\special{em:moveto}}
\put(260,385){\special{em:lineto}}
\put(265,380){\special{em:lineto}}
\put(395,305){\special{em:moveto}}
\put(400,310){\special{em:lineto}}
\put(405,305){\special{em:lineto}}
\put(390,450){\special{em:moveto}}
\put(395,450){\special{em:lineto}}
\put(400,450){\special{em:moveto}}
\put(405,450){\special{em:lineto}}
\put(410,450){\special{em:moveto}}
\put(410,455){\special{em:lineto}}
\put(410,460){\special{em:moveto}}
\put(410,465){\special{em:lineto}}
\put(410,470){\special{em:moveto}}
\put(405,470){\special{em:lineto}}
\put(400,470){\special{em:moveto}}
\put(395,470){\special{em:lineto}}
\put(390,470){\special{em:moveto}}
\put(390,465){\special{em:lineto}}
\put(390,460){\special{em:moveto}}
\put(390,455){\special{em:lineto}}
\put(350,450){\special{em:moveto}}
\put(355,450){\special{em:lineto}}
\put(360,450){\special{em:moveto}}
\put(365,450){\special{em:lineto}}
\put(370,450){\special{em:moveto}}
\put(370,455){\special{em:lineto}}
\put(370,460){\special{em:moveto}}
\put(370,465){\special{em:lineto}}
\put(370,470){\special{em:moveto}}
\put(365,470){\special{em:lineto}}
\put(360,470){\special{em:moveto}}
\put(355,470){\special{em:lineto}}
\put(350,470){\special{em:moveto}}
\put(350,465){\special{em:lineto}}
\put(350,460){\special{em:moveto}}
\put(350,455){\special{em:lineto}}
\put(150,220){\special{em:moveto}}
\put(170,220){\special{em:lineto}}
\put(170,225){\special{em:lineto}}
\put(150,225){\special{em:lineto}}
\put(150,220){\special{em:lineto}}
\put(350,430){\special{em:moveto}}
\put(370,430){\special{em:lineto}}
\put(370,435){\special{em:lineto}}
\put(350,435){\special{em:lineto}}
\put(350,430){\special{em:lineto}}
\put(390,430){\special{em:moveto}}
\put(410,430){\special{em:lineto}}
\put(410,435){\special{em:lineto}}
\put(390,435){\special{em:lineto}}
\put(390,430){\special{em:lineto}}

\put(317,280){\makebox(0,0){\scriptsize $\theta_{\rm II}$}}
\put(330,322){\makebox(0,0){\scriptsize $e^{i\zeta_{\rm II}}$}}
\put(430,362){\makebox(0,0){\scriptsize $e^{i\xi_{\rm II}}$}}
\put(242,240){\makebox(0,0){\scriptsize $U_{\rm II}$}}
\put(237,302){\makebox(0,0){\scriptsize $\phi_{\rm II}$}}
\put(338,240){\makebox(0,0){\scriptsize $\pi \over 2$}}
\put(278,420){\makebox(0,0){\scriptsize $\pi \over 2$}}
\put(385,282){\makebox(0,0){\scriptsize $-{\pi \over 2}$}}
\put(420,332){\makebox(0,0){\scriptsize $\pi$}}
\put(332,420){\makebox(0,0){\scriptsize $\pi$}}
\put(180,220){\makebox(0,0){\scriptsize $V_1$}}
\put(370,420){\makebox(0,0){\scriptsize $V_2$}}
\put(420,430){\makebox(0,0){\scriptsize $V_3$}}
\put(15,50){\makebox(0,0){\scriptsize $\rho$}}
\put(160,250){\makebox(0,0){\scriptsize $E_1$}}
\put(360,460){\makebox(0,0){\scriptsize $E_2$}}
\put(400,460){\makebox(0,0){\scriptsize $E_3$}}
\end{picture}

\caption{The full setup for three measurement operators, using two modules. Note the phase shifters with phase factors $\zeta_{\rm I}$,$\zeta_{\rm II}$,$\xi_{\rm I}$ and $\xi_{\rm II}$, as well as additional unitary operators $U_I$, $U_{II}$, $V_1$, $V_2$ and $V_3$ introduced in order to make the setup completely general. A general state $\rho$ entering the setup becomes the output state $\rho_i = {M_i \rho M_i^\dagger \over {\rm tr}(M_i \rho M_i^\dagger)}$ at exit $E_i$ with probability $p_i = {\rm tr}(M_i \rho M_i^\dagger)$.}\label{twomod}

\end{figure}

If we wish to perform any POVM consisting of three measurement operators, we implement the first POVM operator as $F_1$ in the first module and redirect the amplitude emerging from the other exit into a second module, which
acts upon it with different initial rotation and different $\theta$ and $\phi$ parameters, implementing the remaining operators $F_2$ and $F_3$, so that:

\begin{eqnarray}\label{F}
F_1 &=& U^{\dagger}_{\rm I} D_{\rm I}^2 U_{\rm I}
\cr
F_2 &=& U^\dagger_{\rm I} \tilde{D}^\dagger_{\rm I} U^\dagger_{\rm II} D_{\rm II}^2 U_{\rm II} \tilde{D}_{\rm I} U_{\rm I} 
\cr
F_3 &=& U^\dagger_{\rm I} \tilde{D}^\dagger_{\rm I} U^\dagger_{\rm II} (I - D_{\rm II}^2) U_{\rm II} \tilde{D}_{\rm I} U_{\rm I} = I - F_1 - F_2  
\end{eqnarray}
 
where $U_{\rm I}$ and $U_{\rm II}$ are the unitary operators implemented on the photon before modules I and II respectively, $D_{\rm I}$ and $D_{\rm II}$ correspond to $D_1$ in the modules I and II respectively, and $\tilde{D}_{\rm I}$ plays the role of $D_2$ so that $\tilde{D}^\dagger_{\rm I} \tilde{D}_{\rm I} = I - D_{\rm I}^2$.
Figure \ref{twomod} shows the complete setup for performing any POVM with three measurement operators. 

\vspace{-0.5cm}

The generalization to $n$ operators, using $n-1$ modules, is straightforward.
In general, for $n$ operators and $n-1$ modules we have, for $j < n$:

\begin{equation}
F_j = \left[\prod_{i = 1}^{j-1} U^\dagger_{(i)} \tilde{D}^\dagger_{(i)}\right]
U^\dagger_{(j)} D^2_{j} U_{(j)}
\left[\prod_{i = 1}^{j-1} \tilde{D}_{(i)} U_{(i)} \right]
\end{equation}

and the last operator ($j = n$) is given by:

\begin{equation}
F_n = \left[\prod_{i = 1}^{n-1} U^\dagger_{(i)} \tilde{D}^\dagger_{(i)}\right]
\left[\prod_{i = 1}^{n-1} \tilde{D}_{(i)} U_{(i)}\right]
\end{equation}

where have introduced the notation $U_{\rm I} = U_{(1)}, U_{\rm II} = U_{(2)}, \dots$ and $U_{\rm I} = U_{(1)}, U_{\rm II} = U_{(2)}, \dots$, etc.
  
Note that any probabilistic POVM on a single photon polarization state can be realized with a statistical mixture of at most four POVM operators. If however our setup is to be used for realizing a fully deterministic $n$-operator POVM, then $n-1$ modules are needed. 

\section{Examples}

As an example of the simplicity of applying this procedure, we implement the non-orthogonal projective POVM with the three axes of projection separated by 120 degrees, as discussed in our previous work \cite{us}. However, in contrast to our previous example, we now do not require any delay lines in order to recycle photons around the apparatus in order to implement these operators with unit probability of success.

If we consider:

\begin{equation}
U_{\rm I} =
\left(
\begin{array}{cc}
1 & 0 \cr 0 & 1
\end{array}
\right)
\,\,\,\,\,
U_{\rm II} =
{1 \over \sqrt{2}}
\left(
\begin{array}{cc}
1 & 1 \cr -1 & 1
\end{array}
\right)
\end{equation}

\begin{equation}
D_{\rm I} =
\sqrt{2 \over 3}
\left(
\begin{array}{cc}
1 & 0 \cr 0 & 0
\end{array}
\right)
\,\,\,\,\,
D_{\rm II} =
\left(
\begin{array}{cc}
1 & 0 \cr 0 & 0
\end{array}
\right)
\end{equation}

which corresponds to $\theta_{\rm I} = \arccos(\sqrt{2 \over 3})$, $\theta_{\rm II} = 0$, $\phi_{\rm I} = {\pi \over 2}$ and $\phi_{\rm II} = {\pi \over 2}$. The angles of inital polarization rotation at modules I and II are $0$ and ${\pi \over 4}$ respectively.

Then using equation (\ref{F}), we find that these give:

\begin{equation}
F_1 =
{2 \over 3}
\left(
\begin{array}{cc}
1 & 0 \cr 0 & 0
\end{array}
\right)
\end{equation}

\begin{equation}
F_2 =
{1 \over 6}
\left(
\begin{array}{cc}
1 & \sqrt{3} \cr \sqrt{3} & 3
\end{array}
\right)
\end{equation}

\begin{equation}
F_3 =
{1 \over 6}
\left(
\begin{array}{cc}
1 & -\sqrt{3} \cr -\sqrt{3} & 3
\end{array}
\right)
\end{equation}

as required for this POVM. Note that if we want our apparatus to output photon states whose polarizations are separated by 120 degrees we can simply implement the appropriate Kraus operators while leaving $\{F_i\}$ unchanged, by choosing

\begin{equation}
V_1 = I 
\,\,\,
V_2 = 
{1 \over 2}
\left(
\begin{array}{cc}
1 & -\sqrt{3} \cr \sqrt{3} & 1
\end{array}
\right)
\,\,\,
V_3 = 
{1 \over 2}
\left(
\begin{array}{cc}
1 & \sqrt{3} \cr -\sqrt{3} & 1
\end{array}
\right)
\end{equation}

A second example which illustrates the flexibility of this approach is the implementation of the POVM employed in the quantum cryptography protocol proposed by Ekert {\em et al.} \cite{peres,ekert,brandt97,brandt99}. The operators can be considered in terms of two polarization states with polarizations at angles $\alpha$ and $\beta$:

\begin{equation}
F_1 =
{1 \over 1+\cos(\beta-\alpha)}
\left(
\begin{array}{cc}
\sin^2 \alpha & -\sin \alpha \cos \alpha \cr -\sin \alpha \cos \alpha & \cos^2 \alpha \cr
\end{array}
\right)
\end{equation}

\begin{equation}
F_2 =
{1 \over 1+\cos(\beta-\alpha)}
\left(
\begin{array}{cc}
\sin^2 \beta & -\sin \beta \cos \beta \cr -\sin \beta \cos \beta & \cos^2 \beta \cr
\end{array}
\right)
\end{equation}

and 

\begin{equation}
F_3 = I - F_1 - F_2
\end{equation}

These can be implemented using:

\begin{equation}
U_{\rm I} =
\left(
\begin{array}{cc}
\cos \alpha & \sin \alpha  \cr -\sin \alpha & \cos \alpha \cr
\end{array}
\right)
\,\,\,\,\,
U_{\rm II} =
\left(
\begin{array}{cc}
\cos \alpha' & \sin \alpha'  \cr -\sin \alpha' & \cos \alpha' \cr
\end{array}
\right)
\end{equation}

where $\alpha' = {\rm arccot}[\sqrt{1 + {1 \over \cos(\beta - \alpha)}} \cot (\beta - \alpha)]$, and

\begin{equation}
D_{\rm I} =
\left(
\begin{array}{cc}
0 & 0 \cr 0 & \sqrt{1 \over 1 + \cos (\beta - \alpha)} \cr
\end{array}
\right)
\,\,\,\,\,
D_{\rm II} =
\left(
\begin{array}{cc}
0 & 0 \cr 0 & 1
\end{array}
\right)
\end{equation}

Note that for projective POVMs (i.e. where the Kraus operators are outer products), one of the elements of $D$ will always be zero, as the set of output states is independent of the input state. Our setup however is more general, and encompasses all POVMs, including non-projective sets, whose output states depend on the input states.

\section{Conclusion}

We have presented a linear optics setup for the deterministic implementation of an arbitrary Positive Operator Value Measurement (POVMs) of single-photon polarization states, with any number of operators. The only ancillas required for this implementation are vacuum states.
Our method is completely general and includes all possible sets of POVM operators, namely projective orthogonal, projective non-orthogonal and non-projective. 

Thus any possible generalized quantum measurement of single photon polarization states can be easily performed using this setup.

Due to its deterministic nature this POVM setup could be used to perform POVMs on members of entangled states with the correct multipartite post-measurement state as given by (\ref{povmstat}). Another interesting question arising from this is to ask what the most general setup for multipartite POVMs could be.  

The authors wish to thank T. Rudolph, G. Pryde, N. Peters, P. G. Kwiat and S. Virmani for helpful discussions.

Sebastian Ahnert was supported by the Howard Research Studentship of Sidney Sussex College, Cambridge.

\end{multicols} 
\end{document}